\documentclass{article}
\begin{document}
\def\beq{\begin{equation}} 
\def\eeq{\end{equation}}
\newcommand{\bpi}{\mbox{\boldmath $\pi$}}
\def\theequation{\arabic{equation}}
\newcommand{\sgn}{\mathop\mathrm{sgn}}
\renewcommand{\r}{r_0}
\newcommand{\ii}{\mathrm{i}}
\begin{center}
{\Large Phases in   noncommutative quantum mechanics 
on (pseudo)sphere}\\
\vspace{0.5 cm}
{\large Stefano Bellucci$^1$ and Armen Nersessian$^{1,2,3}$ }
\end{center}
{\it $^1$ INFN, Laboratori Nazionali di Frascati,
 P.O. Box 13, I-00044, Italy\\
$^2$ Yerevan State University, A. Manoogian St., 3, Yerevan,
375025, Armenia\\
 $^3$JINR, Bogoliubov Laboratory of Theoretical
Physics, 141980 Dubna, Russia}

\begin{abstract}
We compare   the  non-commutative quantum mechanics (NCQM)
 on sphere  and  the discrete part of the spectrum  of NCQM
on pseudosphere (Lobachevsky plane, or $AdS_2$) in the presence 
of a constant  magnetic field $B$ with planar NCQM. We show, that 
(pseudo)spherical NCQM has a ``critical point'', where the system becomes
effectively  one-dimensional,  and  two different `` phases'',
 which the phases of the planar system originate from, specified 
by the sign of the parameter  $\kappa=1-B\theta$.
The ``critical point'' of (pseudo)spherical NCQM  corresponds to the
$\kappa\to\infty $ point of conventional planar NCQM, and to 
the ``critical point''   $\kappa=0$ of the so-called
``exotic'' planar NCQM, with a  symplectic coupling of the
(commutative) magnetic field.
\end{abstract}
\subsection*{Introduction}
Noncommutative quantum field theories have been
 studied intensively during the last several years owing
to their relationship with M-theory compactifications \cite{cds},
 string theory in nontrivial backgrounds \cite{sw} and  
 quantum Hall effect \cite{hall}
(see e.g. \cite{szabo} for a recent review).
 At low energies the one-particle sectors become 
relevant, which prompted an interest in the study of 
noncommutative quantum mechanics (NCQM) \cite{LM}-\cite{pAdS}
(for some earlier studies of NCQM see
\cite{Dunne:1990hv}-\cite{Lukierski:1997br}).
In these studies some  attention was paid 
to  two-dimensional NCQM in the presence of a constant
magnetic field: such systems were considered  on a plane \cite{np,our},
 torus \cite{torus}, sphere \cite{np}, pseudosphere 
(Lobachevsky plane, or $AdS_2$ ) \cite{iengo, pAdS}.

 NCQM on a  plane has a critical point,
 specified by the zero value of the dimensionless parameter 
  \begin{equation}
 \kappa=1-B\theta,
\label{int} \end{equation}
where the system  
becomes effectively one-dimensional \cite{np,our}.
Out of the critical point, the rotational  properties of the model
 become qualitatively dependent on the sign 
of $\kappa$: for $\kappa>0$ the system could have
an infinite number of states with a  given value of the angular momentum, 
while for $\kappa<0$ the number of such states is  finite \cite{our}
(see also \cite{jellal}).
In appropriate limits the NCQM on a (pseudo)sphere
 should be reduced to the planar one. Hence, out
of the (pseudo)spherical system 
  originate,  in some sense, the ``phases'' of planar NCQM.
 Although this issue was touched upon in 
Refs. \cite{np,Karabali:2001te,iengo,pAdS},
 the complete understanding of this question  has yet to be achieved.

In  the present paper, we study the relationship between NCQM on 
(pseudo)sphere and plane.
 Considering the planar limit of NCQM on (pseudo)sphere 
 we are lead to the conclusion that NCQM on a sphere
and  the discrete part of the spectrum 
 of NCQM on pseudosphere possess the ``phases'',
which  yield the `` phases'' of the planar system.
The ``critical point'' of (pseudo)spherical NCQM results
in the point $\kappa\to\infty$ of 
the  planar system suggested in Refs. \cite{np,our},
 and in  the ``critical point''  $\kappa=0$ 
of the so-called ``exotic'' NCQM \cite{Duval:2000xr},
 where the magnetic 
field is introduced via  ``minimal'', or symplectic coupling. 

\subsection*{ NCQM on plane, sphere and pseudosphere}
The  ``conventional'' two-dimensional noncommutative quantum mechanical
system with arbitrary central potential in the presence 
of a constant magnetic
field $B$, suggested by Nair and Polychonakos, 
 is given by the Hamiltonian \cite{np},
 \begin{equation}
 {{\cal H}}^{\rm plane}=\frac{{\bf p}^2}{2}+ V({\bf q}^2), \label{h0}
 \end{equation}
and  the operators ${\bf p}, {\bf q}$ which obey the commutation relations
 \begin{equation}\label{xp}
 [q_1,q_2]=i\theta,\quad [{ q}_\alpha, 
{ p}_\beta]=i \delta_{\alpha\beta},\quad
 [p_1,p_2]=iB,\quad\quad\alpha,\beta=1,2
\end{equation}
where the noncommutativity parameter  $\theta>0$ 
has the dimension of ${\rm{\it length}}^2$.

There exists a so-called ``exotic'' NCQM suggested  by 
Duval and Horvathy \cite{Duval:2000xr}. 
Its difference from the ``conventional'' planar 
NCQM lies in the coupling of external magnetic field.
 Instead of a naive, or algebraic approach, 
used in conventional NCQM, the minimal,
 or symplectic, coupling is used there,
 in the  spirit of Souriau \cite{sour}.
This  coupling  assumes  that the closed two-form 
describing the magnetic field
is added 
 to the symplectic structure
 of the underlying Hamiltonian mechanics 
\begin{equation}
\left({\cal H}^{plane}, \omega_0=\theta
dp_1\wedge dp_2 +d{\bf q}\wedge d{\bf p}\right)\to\left({\cal H}^{plane},
 \omega_0+ Bdq_1\wedge dq_2 \right).
\end{equation}
The corresponding 
quantum-mechanical commutators (out of the point $\kappa=0)$ read
\begin{equation}\label{xpe}
 [q_1,q_2]=\ii\frac{\theta}{\kappa}
,\quad [{ q}_\alpha , { p}_\beta ]=\ii\frac{
\delta_{\alpha\beta}}{\kappa}
,\quad [p_1,p_2]=\ii\frac{B}{\kappa}.
\end{equation}
The Hamiltonian is the same as in the ``conventional''  NCQM, (\ref{h0}).

It is convenient to represent these systems as follows:
 \begin{equation}
 {\cal H}^{plane}=\frac{(\bpi+ {\bf q}/\theta )^2}{2} +  V({\bf q}^2),
 \label{8}\end{equation}
where  the operators ${\bpi}$ and ${\bf q}$ are given by the expressions  
 \begin{equation}
{\pi}_1 =p_2-\frac{q_1}{\theta};\;
-{\pi}_2 =p_1+\frac{q_2}{\theta}:[{pi}_\alpha,{ q}_\beta ] = 0,\
\left\{
\begin{array}{cc}
[\pi_1,\pi_2]=-\ii{\kappa}/\theta, \; [q_1,q_2]=i\theta.
&{\rm conventional} \cr 
[\pi_1,\pi_2]=-{\ii}/{\theta}, \; [q_1,q_2]=
\ii{\theta}/{\kappa},
&{\rm exotic}
\end{array}\right. 
\label{7}\end{equation}
The  angular momentum of these systems is defined by the operator
(out of the   point $\kappa=0$)
\begin{equation}
L=\left\{\begin{array}{cc}
{{\bf q}^2}/{2\theta}-{\theta {\bpi^2}}/{2\kappa}& {\rm conventional}\\
\kappa{{\bf q}^2}/{2\theta}-{\theta {\bpi^2}}/{2}&{\rm exotic}
\end{array}\right.
\label{ang}\end{equation}
Its  eigenvalues are given by the expression
\beq
l=\pm(n_1-\sgn\kappa\; n_2),\quad n_1, n_2=0, 1,..
\label{amp}\eeq
where $(n_1, n_2)$ define, respectively,  the eigenvalues of
the operators
$({\bf q}^2,{\bpi}^2)$ for the ``conventional'' NCQM
and of the $({\bpi}^2,{\bf q}^2)$ for the ``exotic'' one, 
the upper sign corresponds to the ``conventional'' system, 
and the lower sign to the ``exotic'' one.
Hence, the rotational properties of NCQM qualitatively depend
on the sign of $\kappa$.
 
At the  ``critical point'', i.e. for $\kappa=0$, these systems
 becomes effectively
one-dimensional \cite{our,Duval:2000xr}
\begin{equation}
[q_1,q_2]=i\theta,\quad {\cal H}^{plane}_0=\left\{
\begin{array}{cc}
{{\bf q}^2}/{2\theta^2}+{ V}({\bf q}^2),&{\rm conventional}\cr
   { V}({\bf q}^2),&{\rm exotic}
\end{array}\right. .
 \label{E0}\end{equation}
Let us remind \cite{Duval:2000xr}, that for non-constant  $B$ 
 the Jacobi identities failed in the ``conventional'' model, while 
in the ``exotic'' model the Jacobi identities hold for any 
$B=A_{[1,2]}$, by definition.
This reflects the different origin of magnetic fields  $B$ appearing
in these two models. In the ``conventional'' model,  $B$ appears
as the strength of a {\it non-commutative } magnetic field,
 while in the  ``exotic'' model, $B$ appears as
a {\it commutative} magnetic
field, obtained by the Seiberg-Witten map from the
non-commutative one. In the quantum-mechanical context
this question was considered in \cite{LM}.\\

The Hamiltonian of   the  
axially-symmetric NCQM on the sphere 
\cite{np,Karabali:2001te,der} 
and pseudosphere  \cite{iengo,pAdS} in the presence 
of a constant magnetic field, looks precisely 
as in the commutative case (up to the dimensionless
parameter $\gamma$)
\footnote{In \cite{np,iengo} the constant term 
 $\mp\gamma s^2/2\r^2$  was ignored; also
 in  \cite{iengo}   the  factor $\gamma$ was
 chosen  to be unity as well.}
\begin{equation} {\cal H}=\pm\gamma\frac{J^2-s^2}{2\r^2}
+ V({\bf x}^2),
\label{hsph} \end{equation}
where the rotation and position operators ${ J}_i=({\bf J}, J_3)$,  
${ x}_i=({\bf x}, x_3)$ obey commutation relations
 \begin{equation} [J_i, J_j]=i\epsilon_{ijk}J^k , \quad
 [J_i,x_j]=i\epsilon_{ijk}x^k ,\quad
 [x_i,x_j]=i\lambda\epsilon_{ijk}x^k,\quad\quad i,j,k=1,2,3.
 \label{algebra}\end{equation} 
Here and after, for squaring the operators and for rising/lowing the
indices, we use the  diagonal metric $diag\; (1,1,1)$ for the sphere 
and $ diag\;(-1,-1,1)$ for the pseudosphere. The upper sign corresponds to 
 a sphere, and the lower one to a pseudosphere. 
The noncommutativity parameter $\lambda$ has the dimension of
{\it length} and is  assumed to be positive, $\lambda>0$.
The values of the Casimir operators of
the algebra are fixed  by the equations
\begin{equation}\label{casimirs}
  C_0\equiv x^2=\r^2>0,\qquad C_1\equiv {J}{x}-
\frac{\lambda J^2}{2}=-\r S(s,\r) ,
\end{equation}
where $\r$ is the radius of the (pseudo)sphere and 
$s$ is the ``monopole number''.
 In the 
commutative limit $\lambda\to 0$ the parameters $S$  
and  $\gamma$
should have  a limit
\begin{equation}\label{mf}
\lambda\to 0\;\Rightarrow \gamma\to 1,\quad S(s,\r)\to s=B{\r^2},
\end{equation}
where $B$ is a strength of the magnetic field.\\ 
The angular momentum of the system is defined by the operator $J_3$:
$[{\cal H}, J_3]=0 $.

The  algebra (\ref{algebra}) can be split in two independent copies
 of su(2)/su(1.1),
\begin{equation}\label{split}
  {K}_i={J}_i-\frac{x_i}{\lambda}:\qquad
  [K_i,x_j]=0,\qquad [K_i,K_j]=\ii \epsilon_{ijk}K^k,
\qquad [x_i,x_j]=i\lambda\epsilon_{ijk}x^k.
\end{equation}
In these terms the   Casimir operators read $C_0=x^2$
and $C_1=\lambda (x^2-K^2)/2 $.
 For the NCQM on  a sphere, the Casimir operators $C_0$, $K^2$
are positive.
For  a pseudosphere  $C_1$ is positive, whereas
another Casimir operator, i.e. $K^2$, could get positive,
zero or  negative
values. We restrict ourselves to the case of positive $K^2$
 which is  responsible for the description 
of the discrete part  of the  energy spectrum. 
Hence, the Casimir operators take the following values:
\begin{equation}\label{c0}
  {\r^2}={\lambda^2}m(m\pm 1),\quad 2s\r+\dots ={\lambda}[k(k\pm 1)-m(m\pm 1)],
\end{equation}
where  $m, k$ are non-negative  (half)integers  fixing the representation  
of SU(2), in the case of sphere, and  $m,k>1$ are   real numbers, 
fixing the representation of SU(1.1), in the case of pseudosphere.

It is unclear, how the qualitatively different planar ``phases''
$\kappa>0$ and $\kappa<0$ originate in the (pseudo)spherical NCQM,
as well as whether the limit of (pseudo)spherical NCQM
results in the ``conventional'' or the ``exotic'' planar
system.
Some steps in relating (pseudo)
spherical NCQM with ``conventional'' planar system
 were performed in  \cite{np,Karabali:2001te,pAdS} .
In particular,  the following expressions for $\gamma$, $\kappa$, $s$ 
parameters were found there:
\begin{equation}
\kappa\approx\pm k/m, \quad \gamma=\kappa,\quad s=k-m,
\label{pparam}\end{equation}
while the arising of planar ``phases''
was explained as a projection ``one to two''(?).
Another  inconsistency of the above picture is that upon  the choice of
 parameters (\ref{pparam}),  the (pseudo)spherical NCQM becomes 
effectively one-dimensional at the point ${\tilde k}=0$,
which  yields  $\kappa\to\infty$, instead of $\kappa=0$.
At this point,  the Hamiltonian consists of a  potential term only,
which seems to be in agreement
 with the planar ``exotic'' Hamiltonian at the point $\kappa=0$.

In order to clarify the above listed questions,
 in the next section we compare 
the planar limits of NCQM on sphere and  of the  discrete part
 of the NCQM on pseudosphere  with both ``conventional'' and ``exotic''
versions  of planar NCQM.
 
\subsection*{NCQM: (pseudo)sphere $\to$ plane}
In order to obtain  the  planar limit of  the NCQM 
on the (pseudo)sphere out of the point $\kappa=0,$ we should 
take the limits \cite{np}
\begin{equation}
k\to \infty, \quad m\to\infty,
\label{30}\end{equation}
and consider  small neighborhoods  of the ``poles'' of
``coordinate and momentum spheres'' 
\begin{equation} 
x_3\approx \epsilon_1(\r\mp\frac{{\bf x}^2}{2\r})=
\epsilon_1\lambda\left({\tilde m}\mp
\frac{{\bf x}^2}{2\lambda^2{\tilde m}}\right),\quad
k_3\approx\epsilon_2 \left({\tilde k}\mp\frac{{\bf K}^2}{2{\tilde k}}\right),
 \qquad\epsilon_{1},\epsilon_2=\pm 1 .
\label{31}\end{equation}
In these  neighborhoods   the commutation
relations 
\begin{equation}
[x_1,x_2]\approx \ii\epsilon_1\lambda^2{\tilde m},
\quad
 [K_1,K_2]\approx
 \ii\epsilon_2{\tilde k}
\label{32}\end{equation}
hold, 
while  the Hamiltonian looks as follows:
\begin{equation}
  {\cal H}=\pm\gamma\frac{
{\tilde k}^2\pm 2{\bf x}{\bf K}/{\lambda}+2k_3x_3/\lambda
+{\tilde m}^2-s^2}{2\r^2}+ V({\bf x}^2)
\approx {\cal E}_0
-\epsilon
 \gamma\frac{ (\nu{\bf K}-
\epsilon {\bf x}/\lambda\nu)^2}{2\r^2}+ V({\bf x}^2).
\label{33}\end{equation}
Here we  introduced the notation
$${\tilde m}=\sqrt{m(m\pm1)},\quad{\tilde k}=\sqrt{k(k\pm1)}\quad
\nu=\sqrt{{\tilde m}/{\tilde k}},\quad
\epsilon=\epsilon_1\epsilon_2$$
and
\begin{equation}
{\cal E}_0= \pm\gamma\frac{
({\tilde k}+\epsilon {\tilde m})^2-s^2 }{2\r} .
\label{e0}\end{equation}
In order to get the  planar Hamiltonian with a positively 
defined kinetic  term, we should  put
\begin{equation}
\sgn \gamma =-\epsilon
\label{spole}
\end{equation} 
For a correspondence 
 with the  planar  Hamiltonian (\ref{8}), 
 we  redefine  the coordinates and momenta 
of the resulting system  as follows:
\begin{equation}
  {\bpi}=\frac{\sqrt{|\gamma|}\nu {\bf K}}{\r},\quad
  \frac{{\bf q}}{\theta}=\frac{\sqrt{|\gamma|}{\bf x}}{\nu\lambda\r}.
\label{34}\end{equation} 
Then,  comparing their commutators with  (\ref{7}), we get 
the  following expressions for  the  $\theta$ 
 parameters:
\begin{equation}
{\theta}=
\left\{\begin{array}{cc}
{\lambda^2{\tilde m}^2}/{\gamma{\tilde k}},
&{\rm conventional} \\
{\lambda^2}{\tilde m}/\gamma,
&{\rm exotic }
\end{array}\right.
\label{spartheta}\end{equation} 
and the same value of $\kappa$ for both systems
\begin{equation}
\kappa=-\epsilon\frac{\tilde m}{\tilde k}.
\label{spar}\end{equation}
Naively,  it seems that the planar NCQM with $\kappa<0$
and positive kinetic term corresponds to  a (pseudo)spherical system  
with negative kinetic term.
Fortunately, thanks to the  additional term $-\gamma s^2/2\r^2$ the 
kinetic term  of the Hamiltonian (\ref{hsph}) remains positively defined!
Indeed, one can identify the monopole number $s$ as follows:
\begin{equation}
 s=\left\{
\begin{array}{cc}
{\tilde m}+\epsilon {\tilde k},& {\rm conventional}\\
-({\tilde m}+\epsilon {\tilde k}),& {\rm exotic}\\
\end{array}\right.   
 \label{s}\end{equation}
which yields
 the vanishing of the ``vacuum energy'' (\ref{e0}), 
and the following expressions  for the magnetic field,
 which are in agreement with (\ref{mf}):
 \begin{equation}
{\tilde  B}=\frac{B}{1-B\theta}=\frac{1-\kappa}{\theta\kappa}=
\left\{
\begin{array}{cc}
\gamma {s}/\kappa{\r^2},&{\rm conventional} \\
\gamma {s}/{\r^2},&{\rm exotic} 
\end{array}\right. .
\label{mfs}\end{equation}
One can   redefine the parameters $s$, $\kappa$  as follows:
\begin{equation}
\kappa=-\epsilon\frac{m\pm 1/2}{k\pm 1/2},\quad
 s=\left\{\begin{array}{cc}
k\pm 1/2+\epsilon (m\pm 1/2),&{\rm conventional}\\
- (k\pm 1/2)-\epsilon (m\pm 1/2),&{\rm exotic}
\end{array}\right. .
\label{sq}\end{equation} 
In this case  the   
 monopole number is quantized on  a sphere, and
it remains not quantized 
 on  a pseudosphere, as in the commutative case.
 The constant energy term ${\cal E}_0$ vanishes upon this 
choice too.

Taking into account that the
maximal value of $J^2$  is $(k+m)(k+m\pm 1)$, 
 and the minimal one is $|k-m|(|k-m|\pm 1)$ \cite{ll}, we obtain 
\begin{equation}
\pm\epsilon\frac{J^2-s^2}{2\r^2}\geq 0.
\end{equation}
Hence, the kinetic  part of the (pseudo)spherical Hamiltonian 
is positively defined for any  $\gamma$.
Expanding (pseudo)spherical NCQM  near the  
upper/lower  bound of $J^2$, we shall  get the planar  NCQM 
with $\kappa>0$/$\kappa<0$.  

 In order to avoid the rescaling of the potential in the 
 planar limit, we should take
\begin{equation}
\gamma=\left\{
\begin{array}{cc}
\kappa,\quad \Rightarrow \lambda={\theta}/{\r}
&{\rm conventional} \\
{1}/{\kappa}\quad
 \Rightarrow \lambda={\theta}/{\kappa\r}
&{\rm exotic} \\
\end{array}\right. .
\label{gammas}\end{equation}
Upon this choice, the expression (\ref{mfs}) reads
\begin{equation}
\frac{s}{\r^2}=\left\{
\begin{array}{cc}
 { \tilde B} &{\rm conventional} \\
 { B} &{\rm exotic}
\end{array}\right. .\label{mfsw}
\end{equation}
In the ``conventional'' picture   ${\tilde B}$ 
plays  the role of  the strength of a
(commutative) magnetic field 
obtained by the Seiberg-Witten map from the
non-commutative one \cite{np}.
In the ``exotic'' picture the same role is played by $B$.
Hence, in both  pictures we
get the standard expression for the 
strength of the constant commutative magnetic field on  (pseudo)sphere,
and the quantization of the
flux of the commutative magnetic field  
on the sphere, as well.

We did not consider yet the planar limit of 
the critical point of (pseudo)spherical
NCQM, and  did not establish yet, whether the latter
results in the ``conventional''
or in the ``exotic'' planar NCQM, in this
limit.
For this purpose let us notice, that 
our specification of the ``monopole number'' $s$ and of the  $\gamma$
 parameter  yields the following 
values of  the  first   Casimir operator:
\begin{equation}\label{fcasimirs}
C_0=\r^2=\lambda^2{\tilde m}^2\;\Rightarrow  \r^2=
\left\{
\begin{array}{cc}
 \theta{\tilde m} &{\rm conventional} \\
 \theta {\tilde k} &{\rm exotic}
\end{array}\right. .
\label{c0f}\end{equation}
Thus, in the ``conventional'' picture the (pseudo)spherical NCQM becomes
one-dimensional for ${\tilde k}=0$, i.e. for $\kappa\to\infty$;
in the ``exotic'' picture we have, instead, ${\tilde m}=0$, i.e. $\kappa=0$.

In the ``exotic'' picture the (pseudo)spherical NCQM in the
$\kappa\to 0$ limit results  in the system
\begin{equation}
{\cal H}_0= V({\bf x}^2),\quad [x_1, x_2]=\ii\theta\sqrt
{1\pm {\bf x}^2/\r^2},
\end{equation}
which  reduces, immediately, to the ``exotic'' planar 
NCQM at the critical point.

Hence, the ``critical point'' and ``phases'' of
(pseudo)spherical NCQM reduce, in the planar limit, 
to the respective ``critical point'' and ``phases'' of
``exotic'' NCQM,  with the symplectic coupling 
of the (commutative) magnetic field.
 
The eigenvalues of the angular 
momentum of (pseudo)spherical NCQM are given by the expression
\begin{equation}
j_3=k_3+m_3, 
\left\{
\begin{array}{ccc}
 k_3 = 0,
\pm 1,\ldots, \pm k,& m_3=0, \pm 1/2,\ldots \pm m &{\rm sphere}\cr
k_3 = \pm k,\pm(k+1),\ldots  
& m_3=\pm m,\pm (m+1),\ldots  & {\rm pseudosphere}
\end{array}\right. .
\label{?}\end{equation}
Introducing $m_3=\epsilon_1(m\mp n_1)$, $k_3=\epsilon_1(k\mp n_2)$,
we get
\begin{equation}
j_3=\epsilon_1(m\mp n_1)+\epsilon_2(k\mp n_2)=
\epsilon_1 \left((m+\epsilon k) \mp (n_1+\epsilon m_2)\right),
\end{equation}
which is in agreement with the angular momentum of 
planar  NCQM (\ref{amp}).
  

\subsection*{Conclusion}
We considered non-commutative quantum mechanics 
on sphere and on pseudosphere  in the presence of constant magnetic field 
(with the strength $B$), and  compared these systems 
 with ``conventional'' \cite{np} 
and ``exotic'' \cite{Duval:2000xr} models 
 of noncommutative  quantum mechanics on plane,
specified  by a different coupling of the
magnetic field.

We have shown that the  quantum mechanics on sphere
and the discrete part of the spectrum  of quantum 
 mechanics on pseudosphere,
defined by the Hamiltonian (\ref{hsph}) and the commutation 
relations (\ref{algebra})
essentially depend on the  values of the Casimir operators, 
\beq
C_0=x^2=\r^2>0,\quad C_1=Jx-\frac{\lambda}{2}J^2=-s\r-\frac{\lambda s^2}{2}\;:
\;C_0\geq\frac{\lambda}{2}C_1,
\eeq
where $\lambda>0$,  $\r$, $s$ are 
 the noncommutativity parameter, the radius
 of sphere and  the monopole number, respectively.

When $C_0={\lambda}C_1/{2}$  (pseudo)spherical NCQM becomes
 effectively one-dimensional.
It yields the ``conventional'' planar NCQM
for $\kappa\to\infty$  and the ``exotic''
one at the point $\kappa=0$,
with the symplectic (or minimal) coupling of the
commutative magnetic field $B$.. 

When $C_0>{\lambda}C_1/{2}$, the monopole number $s$ is connected with 
$r,\lambda$  by the quadratic equation, which has two  solutions, 
corresponding to the positive and negative values of
the specific  parameter $\kappa$.
In the planar limit the ``phases'' of  (pseudo)spherical NCQM 
lead to   the respective ``phases'' of ``conventional'' and
``exotic'' planar NCQM.

\subsection*{Acknowledgments.} 
The authors thank Corneliu Sochichiu for numerous  discussions and
collaboration at the early stages of this research, Evgeny Ivanov for 
clarifying  comments and Philippe Pouliot for his interest in this work.
S.B. is supported in part by the European Community's Human Potential
Programme under contract HPRN-CT-2000-00131 Quantum Spacetime,
the INTAS-00-0254 grant and the Iniziativa
Specifica MI12 of the Commissione IV of INFN. The work of 
 A.N.  is 
supported in part by  the   INTAS  00-00262 and ANSEF  PS124-01 grants.

\end{document}